# Out-of-plane longitudinal sound velocity in $SnS_2$ determined via broadband time-domain Brillouin scattering

Meixin Cheng[1], Kostyantyn Pichugin[1], Andre Maas[2], Marika Schleberger[2] and Germán Sciaini[1*]

[1]The Ultrafast electron Imaging Lab, Department of Chemistry, and Waterloo Institute for Nanotechnology, University of Waterloo, Waterloo, N2L 3G1, Canada.

[2]University of Duisburg-Essen, Faculty of Physics and CENIDE, 47057, Duisburg, Germany.

[*]Correspondence: gsciaini@uwaterloo.ca.

**ABSTRACT**

Here we report time-resolved broadband transient reflectivity measurements performed in a single crystal of SnS$_2$. We made use of time-domain Brillouin scattering and a broadband probe to measure the out-of-plane longitudinal sound velocity, $v_L$ = (2950 ± 100) m s$^{-1}$, in this semiconducting two-dimensional transition metal dichalcogenide. Our study illustrates the potential of this non-invasive all-optical pump-probe technique for the study of the elastic properties of transparent brittle materials and provides the value of the elastic constant $c_{33}$ = (39 ± 3) GPa.

**Keywords:** Impulsive stimulated Brillouin scattering, coherent acoustic phonons, broadband transient spectroscopy, elastic constant, sound speed, sound velocity, transition metal dichalcogenides, 2D materials, SnS$_2$.

## I. INTRODUCTION

The advent of intense and ultrashort laser pulses[1] and the development of different pump-probe schemes have made possible the generation and detection of coherent acoustic phonons approaching the terahertz (THz) range in both absorptive and transmissive solid-state materials[2–13]. The impulsive absorption of photons at the surface of a crystal ultimately results in thermal expansion and the generation of an elastic or acoustic wave that moves away from the surface into the bulk at the speed of sound. The propagation of such a pressure wave leads to time-dependent local density fluctuations, i.e., local changes in the dielectric constant, which in the case of transparent media, can be monitored deep inside the material by a weaker optical probe pulse. The scattering of light due to spontaneous local density variations was first predicted by Léon Brillouin in 1922[14]. Spontaneous Brillouin scattering (BS) is, however, negligible in comparison to the large scattering efficiencies that arise from the dielectric changes induced by the mere presence of a strong electric field – a process that is often referred to as electrostriction and stimulated BS (SBS)[15,16]. Such a phenomenon has been exploited using a single short excitation pulse[17] or two crossing excitation pulses[18]. The latter approach forms the base of impulsive SBS microscopy[19–21], which relies on the generation of density grating that propagates in both directions at the speed of sound of the material and induces a time-dependent modulation on the diffracted probe beam from the acoustic frequency and then the sound speed can be obtained.

In addition, the absorption of an intense ultrashort laser pulse over the optical penetration length at the material's surface leads to impulsive thermal expansion and the creation of a strain pulse that propagates from the surface into the bulk at the sound velocity. Such a strain pulse modulates the dielectric properties of the medium and effectively acts as a reflective interface for a light beam to undergo BS, see Fig 1(a). Such a phenomenon and its geometrical variations gave birth to the technique known as picoacoustics[3–5,7,22], which has been also termed time-domain BS (TDBS)[23,24].

In this work we performed broadband time-domain BS (bb-TDBS) measurements to determine the out-of-plane longitudinal (i.e., perpendicular to the plane of the layers and along the c-axis) sound velocity, $v_L$, in the two-dimensional (2D) transition metal dichalcogenide (TMDC), $SnS_2$. $SnS_2$ has recently attracted great interest owing to its semiconductive character that confers great potential for its application in dye-sensitized solar cells[25], flexible and fast photodetectors[26,27], and field effect transistors[28–30]. However, despite the large body of structural and electronic studies carried out in $SnS_2$[31–35], there is scarce experimental information in regards to its acoustic or elastic properties[36–41]. The latter fact is likely because of its low Pugh or B/G ratio that makes $SnS_2$ quite brittle[37] (B and G correspond to the bulk and shear elastic moduli, respectively). Therefore, we decided to implement an all-optical non-disruptive approach to determine $v_L$ in this material. For a recent review article that includes time-domain acoustic measurements of $v_L$ in different 2D-TMDCs the reader may refer to citation [42].

## II. EXPERIMENTAL METHODS

The sample, a single crystal of $2H$-$SnS_2$, was purchased from HQ Graphene. Note that the usual structure of natural of $SnS_2$ (the mineral berndtite) has been historically referred to as $2H$ with '2' indicating the number of S layers within the primitive unit cell and '$H$' to hexagonal symmetry[43]. However, according to more recent nomenclature rules for transition metal dichalcogenides[44], $SnS_2$ has a crystalline structure that corresponds to $1T$[33], with '1' representing the number of S-Sn-S slabs within the primitive unit cell and '$T$' owing to its trigonal lattice structure. Therefore, the reader may find in literature the same semiconducting crystalline phase of $SnS_2$ labeled as $1T$ or 2$H$, being 2$H$ the most common. We will then omit the use of the prefixes '2$H$' or '1$T$' in most of the remaining text to avoid creating confusion. Figure 1(b) shows the structure of SnS$_2$ and the trigonal primitive unit cell. Bulk 2$H$-SnS$_2$ belongs to the symmetry group of $P\bar{3}m1$ and has a trigonal lattice structure with constants $a = b = 0.36486$ nm and $c = 0.58992$ nm[43]. The single crystal was characterized via Raman spectroscopy. The Raman spectrum showed the strong characteristic A$_{1g}$ ≈ 315 cm$^{-1}$ mode of the 2$H$ phase[35,45] of SnS$_2$ while the XPS analysis revealed that the SnS$_2$ crystal consisted of Sn$^{4+}$ and S$^{2-}$ valence states.

Our bb-TDBS experiments were conducted with a Light Conversion Pharos-SP laser system that delivers 170-fs optical pulses with a central wavelength of 1030 nm at a repetition rate of 6 kHz. The SnS$_2$ crystal was excited by pump pulses with a wavelength, $\lambda_{pump}$ = 400 nm. Pump pulses were generated via optical parametric amplification followed by frequency doubling. The pump beam was focused to a spot size of ≈ 300 μm fwhm (full-width at half-maximum) with an approximate incident fluence of 0.5 mJ·cm$^{-2}$. SnS$_2$ is stable under ambient conditions and has an indirect band gap of ≈ 2.3 eV and direct band gap at room temperature of ≈ 2.4 eV[34] ($\lambda$ ≈ 520 nm). Pumping above the band gap removed the need to introduce a commonly employed transducing metallic layer. The transient reflectivity changes were monitored by broadband white light supercontinuum probe pulses, which covered the photon wavelength range of $\lambda_{probe}$ ≈ 550 nm – 900 nm. Broadband probe pulses were generated by focusing a small fraction of the fundamental beam into a 3-mm thick YAG crystal. The size of the probe beam at the sample was ≈ 50 μm (fwhm). Broadband femtosecond transient reflectivity (bb-TR) measurements were performed at room temperature, 295 K, at three different incident probe beam angles, $\theta$ = $\{15°, 30°, 45°\}$. The relative time between the pump and probe pulses was controlled by an optical delay stage placed in the probe beam path. The broadband reflectivity spectra were recorded by a dispersive spectrometer with a frame transfer rate of 1 kHz. The differential transient reflectivity spectra were obtained by modulating the pump beam with a synchronized mechanical chopper operating at 500 Hz.

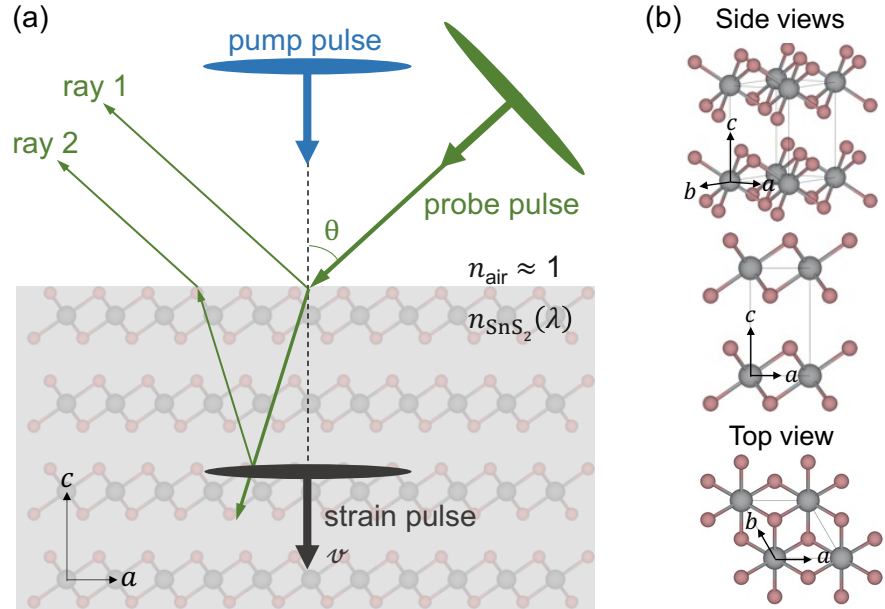

**FIG. 1.** (a) Closer view of the ray propagation geometry. $\theta$ is incident probe beam angle. The pump pulse beam is oriented perpendicularly to the interface, and it is absorbed within the optical penetration-depth at $\lambda_{pump}$ = 400 nm ($\alpha^{-1} = \frac{\lambda_{pump}}{4\pi k} \approx$ 85 nm) from the interface, generating an acoustic pulse that travels into the bulk of the material. This acoustic pulse acts as a moving reflective interface for *ray 2*, which will interfere with *ray 1*, and therefore modulate the detected intensity in the spectrometer depending on the optical path difference and $\lambda_{probe}$. $k$ is the imaginary component of the refractive index and $k(\lambda_{pump}$ = 400 nm) $\approx$ 0.37, black trace in Fig. 3(a). (b) Different views of the crystalline structure of $SnS_2$ and the trigonal primitive unit cell.

Figure 1(a) shows a scheme of our geometrical beam arrangement. The pump beam arrives perpendicular to the surface of the crystal. The probe beam impinges the interface at a selected $\theta$. The difference between the optical paths of *ray 1* and *ray 2* depends on the relative arrival time ($t$) between the probe and the pump pulses to the sample, which is modified by the optical delay stage. Thus, for a given $\lambda_{\text{probe}}$, the detected light intensity will feature a time-dependent modulation because of alternating constructive and destructive interference effects arising from the superposition of the electric fields of *ray 1* and *ray 2* according to the following Eq. 1,

$$2\, v_L\, T\, \sqrt{n^2 - \sin^2\theta} = \lambda_0, \quad (1)$$

Here, $\lambda_0 \equiv \lambda_{\text{probe}}$ in air, $T$ is the period of the observed oscillatory signal in the time domain at a given $\lambda_0$, $v_L$ is the longitudinal sound velocity in SnS$_2$, and $n$ is the real part of its refractive index. Note that $T$ is the inverse of the Brillouin frequency, $v_B$. For more details about our experimental setup and a simple derivation of Eq. 1 based on ray tracing, the reader may refer to reference [46]. We have derived Eq. 1 using the standard beam propagation model that considers only $n$ for ray tracing (direction of the refracted beam) whereas the effect of $k$ (imaginary part of the refractive index) is included as an attenuation factor acting on the amplitude of the refracted beam, i.e., *ray 2*. As a rule of thumb, if $k/n < 0.07$ the standard method works well[47], which is the case for our range of detected $\lambda_0$.

## III. RESULTS AND DISCUSSION

Figure 2(a) shows a raw time-resolved bb-TR spectrum obtained for $\theta \approx 30°$. Two time-dependent TR traces obtained by slicing the spectrum at two different values of $\lambda_{\text{probe}}$ are shown in Fig. 2(b) and 2(c). The main feature associated with the TDBS process is the observed long-lived oscillation, which is superimposed on a slowly varying background that emerges from the dynamics of photoexcited carriers in SnS$_2$. We determined $v_B$ at each $\lambda_0$ by implementing the following fitting model:

$$S(t) = A_1 e^{\frac{-(t-t_1)}{\tau_1}} + A_2 e^{\frac{-(t-t_2)}{\tau_2}} + A_3 e^{\frac{-(t-t_3)}{\tau_3}} \sin(2\pi \nu_B t + \varphi) + C, \qquad (2)$$

where $\tau_{j=1,2,3}$, $A_{j=1,2,3}$, $t_{j=1,2,3}$, $\varphi$, $\nu_B$, and $C$ are fitting parameters; $\tau_{j=1,2,3}$ are time constants, $A_{j=1,2,3}$ are amplitudes, $t_{j=1,2,3}$ are time origins, $\varphi$ is a phase shift with respect to time zero and $C$ is the offset at $t \rightarrow \infty$. Two exponential decays were found to suffice to represent the time-dependent background signal that mostly arises from carrier dynamics and accompanies the oscillatory component of interest. The latter is fitted by a *sine* function (third term in Eq. 2) with a damping factor to account for decoherence effects as well as the weak absorption experienced by *ray 2* at probing wavelengths approaching the band gap of SnS$_2$. Figure 2(d) shows the result obtained from our automated fitting procedure, the similarity between our raw data (Fig. 2(a)) and model calculations (Fig. 2(c)) warrants the proper convergence of our fitting procedure across the complete range of probe wavelengths.

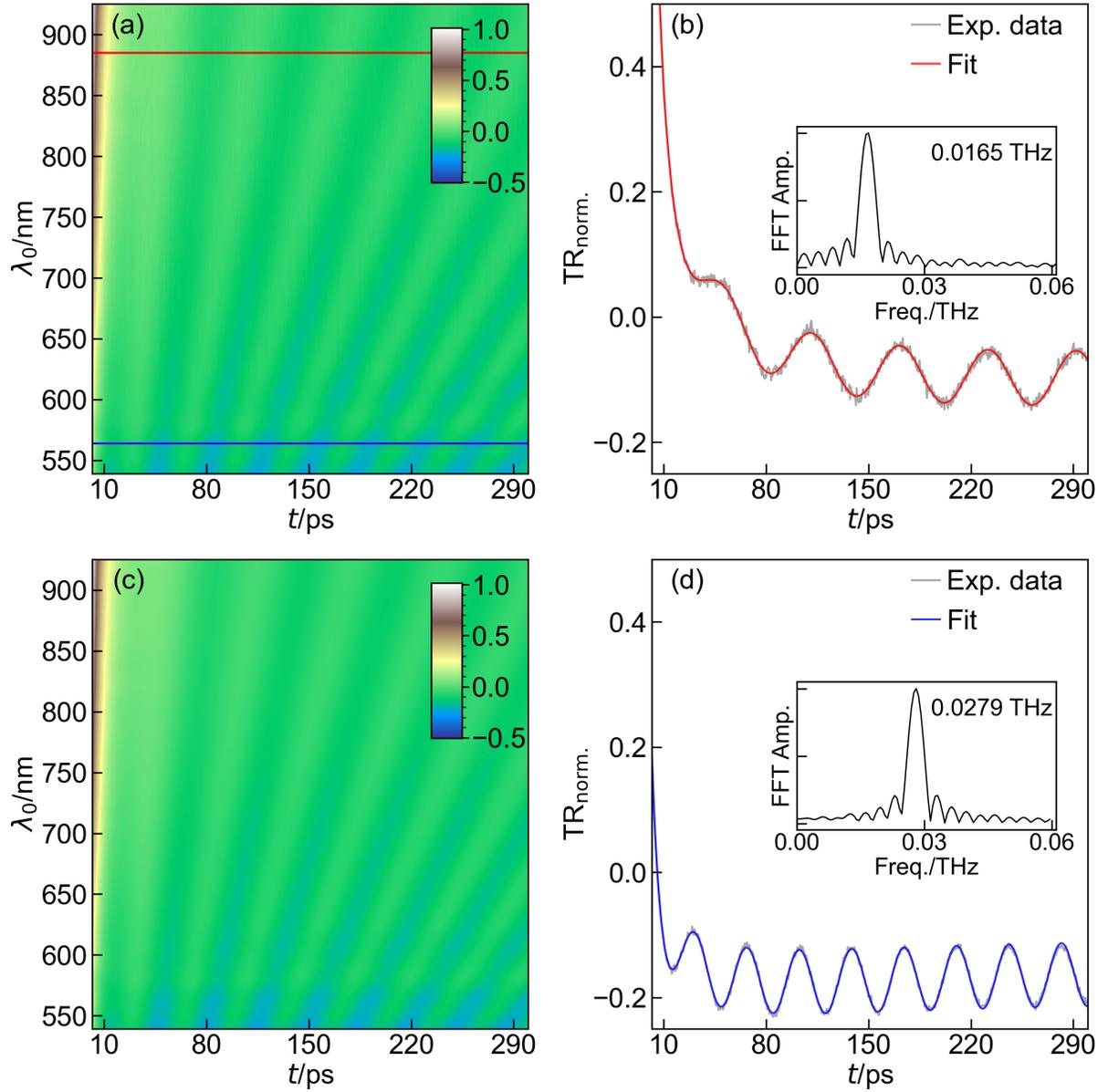

**FIG. 2.** (a) Normalized raw time-resolved bb-TR spectrum collected with a time step of 500 fs. (b, d) The gray traces correspond to slices of the time-resolved bb-TR spectrum at $\lambda_0 \approx 885$ nm (b) and $\lambda_0 \approx 565$ nm (d). The positions of these slices are indicated in panel (a) with the same colors (red and blue) used to show the fitting results implementing Eq. 2. The fast Fourier transform (FFT) spectra of the oscillatory components, attained after removal of the slowly varying background, are shown as insets. (c) Model fitting result obtained by combining all fits as a function of $\lambda_0$ in a single plot to mimic the experimental time-resolved bb-TR spectrum.

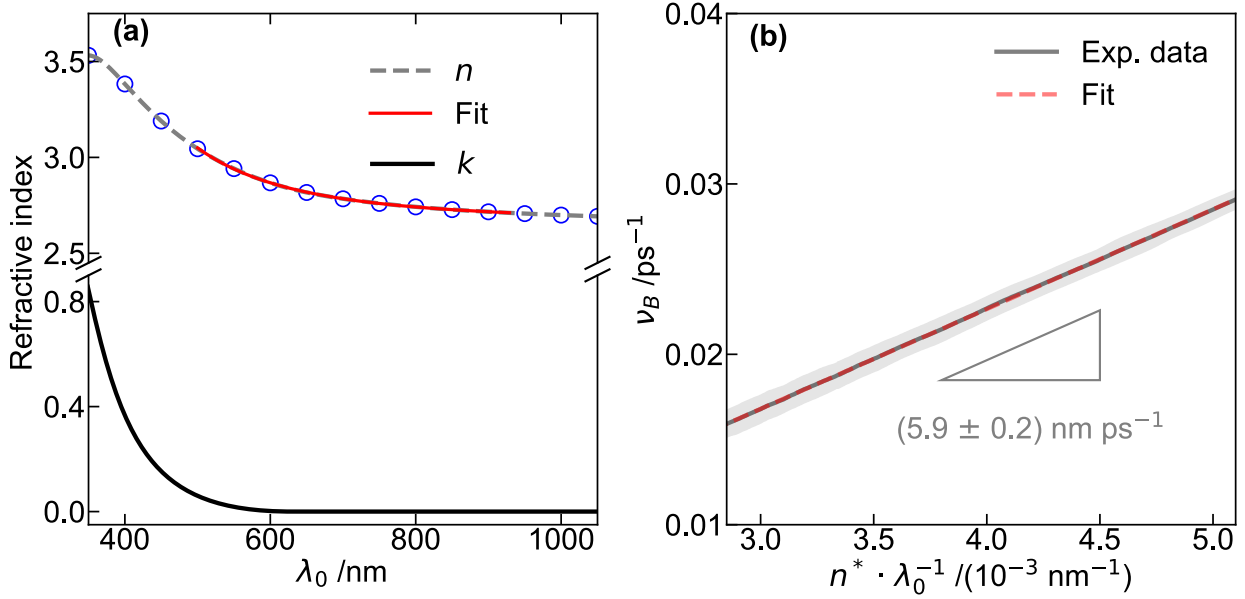

**FIG. 3.** (a) Real ($n$) and imaginary ($k$) parts of the refractive index of SnS$_2$ as a function of wavelength[33]. (b) The dark gray trace corresponds to the plot of $v_B = T^{-1}$ as a function of $n(\lambda_0)^* \cdot \lambda_0^{-1}$ with $n(\lambda_0)^* = \sqrt{n(\lambda_0)^2 - \sin^2(\theta = 30°)}$. The red trace represents a linear fit of the data, which yields a slope of (5.9 ± 0.2) nm ps$^{-1}$ and $v_L$ = (2950 ± 100) km s$^{-1}$. The light gray shadow represents the standard deviation of $v_B(\lambda_0, \theta \approx 30°)$ as obtained from fittings using Eq. 2. The error in $v$ originates mostly from the uncertainty of $\theta$, which was estimated to be ± 5°.

Figure 3(a) presents the values $n$ and $k$ of SnS$_2$ as a function of $\lambda_0$[33]. The values of $n$ were fitted within the wavelength interval of interest using the Cauchy relationship, $n(\lambda_0) = A + B/\lambda_0^2 + C/\lambda_0^4$. The result of this fitting was used as our input in Eq. 1, which can be rearranged as $v_B(\lambda_0) = 2\, v\, n(\lambda_0)^* \lambda_0^{-1}$ (with $n(\lambda_0)^* = \sqrt{n(\lambda_0)^2 - \sin^2\theta}$). Thus, $v_L$ can be calculated from the slope of the linear regression when plotting $v_B(\lambda_0)$ versus $n(\lambda_0)^* \lambda_0^{-1}$ as shown in Fig. 3(b). Note that the values of $n(\lambda_0)$, $\theta$, and $v_B(\lambda_0)$ are known. It should be mentioned the use of a broadband probe is not a must since a single color would suffice to determine $v$. However, broadband probing confers a higher confidence to the value of $v_L$, which is obtained from a linear regression that incorporates the dependence of $n$ with $\lambda_0$. A linear trend also rules out possible calibration issues with the employed dispersive spectrometer[11–13], which is perhaps the only disadvantage with respect to single wavelength detection with a photodiode. We determined a slope of $5.9 \pm 0.2$ nm ps$^{-1}$, which rendered a value for the out-of-plane longitudinal sound velocity of $v_L = 2950 \pm 100$ m s$^{-1}$. Very similar results were obtained from our experiments at $\theta = \{15°, 45°\}$, which provided the values of $v_L = 2900 \pm 100$ m s$^{-1}$ and $v_L = 2950 \pm 100$ m s$^{-1}$, respectively. Our values are in reasonable agreement with that determined indirectly by Mnari et al.[36], $v_L = 3550$ m s$^{-1}$, who had to model acoustic microscopy data owing to substrate effects arising from their SnS$_2$ films deposited on Pyrex glass. Moreover, the observed agreement among the values of $v$ obtained at three different $\theta$ asserts the validity of Eq. 1 and emphasizes the importance of considering the effect of the air-sample interface on the trajectory of *ray 2*[46]. In addition, as expected from the conclusion drawn in our previous study in hematite[46], we found that the combination of measurements at different incident probe angles such that $v_{B,1}^2 - v_{B,2}^2 = 4\, v_L^2 (\sin^2\theta_2 - \sin^2\theta_1)\, \lambda_0^{-2}$ leads to very large error bars, and therefore we decided to omit that data treatment in the current work for clarity. Our value of $v_L = 2950 \pm 100$ m s$^{-1}$ can be related to the elastic constant $c_{33} = \rho\, v_L^2$, with $\rho = 4.46$ g cm$^{-3}$ being the density of SnS$_2$. This provides $c_{33} = (39 \pm 3)$ GPa, which is in reasonable agreement with the calculated value by Zhen and Wang[39], $c_{33} = 32$ GPa. However, it should be mentioned that the values of $c_{33}$, $c_{44}$ and $c_{13}$ depend strongly on the employed method[32,34,36], which highlights the need for experimental values to evaluate the pseudopotentials used in such first-principle DFT calculations.

## IV. CONCLUSIONS

Our study illustrates the relevance of non-destructive techniques to determine the sound velocity, and therefore the elastic properties of brittle two-dimensional layer materials like $SnS_2$, which may not withstand conventional tension and compression tests. Our all-optical method and analysis protocol were found to provide robust results, benefiting from the implementation of a broadband probe.

**Data availability statement:** The data that support the findings of this study are available from the corresponding author upon reasonable request.

**Acknowledgements**

G.S. acknowledges the support of the National Science and Engineering Research Council of Canada, the Canada Foundation for Innovation, and the Ontario Research Foundation. This research was undertaken thanks, in part, to funding from the Canada First Research Excellence Fund. A.M. and M.S. acknowledge financial support from the BMBF (Project 05K16PG1).

## References

[1] D. Strickland and G. Mourou, Opt. Commun. **56**, 219 (1985).

[2] K.A. Nelson, J. Appl. Phys. **53**, 6060 (1982).

[3] C. Thomsen, J. Strait, Z. Vardeny, H.J. Maris, J. Tauc, and J.J. Hauser, Phys. Rev. Lett. **53**, 989 (1984).

[4] O. Matsuda, O.B. Wright, D.H. Hurley, V. Gusev, and K. Shimizu, Phys. Rev. B **77**, 224110 (2008).

[5] A. Devos and R. Côte, Phys. Rev. B **70**, 125208 (2004).

[6] M. Harb, W. Peng, G. Sciaini, C.T. Hebeisen, R. Ernstorfer, M.A. Eriksson, M.G. Lagally, S.G. Kruglik, and R.J.D. Miller, Phys. Rev. B **79**, 094301 (2009).

[7] E. Pontecorvo, M. Ortolani, D. Polli, M. Ferretti, G. Ruocco, G. Cerullo, and T. Scopigno, Appl. Phys. Lett. **98**, 011901 (2011).

[8] P. Ruello, T. Pezeril, S. Avanesyan, G. Vaudel, V. Gusev, I.C. Infante, and B. Dkhil, Appl. Phys. Lett. **100**, 212906 (2012).

[9] S. Ge, X. Liu, X. Qiao, Q. Wang, Z. Xu, J. Qiu, P.-H. Tan, J. Zhao, and D. Sun, Sci. Rep. **4**, 5722 (2014).

[10] N. Rivas, S. Zhong, T. Dekker, M. Cheng, P. Gicala, F. Chen, X. Luo, Y. Sun, A.A. Petruk, K. Pichugin, A.W. Tsen, and G. Sciaini, Appl. Phys. Lett. **115**, 223103 (2019).

[11] S. Brivio, D. Polli, A. Crespi, R. Osellame, G. Cerullo, and R. Bertacco, Appl. Phys. Lett. **98**, 211907 (2011).

[12] A. Devos, Y.-C. Wen, P.-A. Mante, and C.-K. Sun, Appl. Phys. Lett. **100**, 206101 (2012).

[13] S. Brivio, D. Polli, A. Crespi, R. Osellame, G. Cerullo, and R. Bertacco, Appl. Phys. Lett. **100**, 206102 (2012).

[14] L. Brillouin, Ann. Phys. **9**, 88 (1922).

[15] M.J. Damzen and Institute of Physics, *Stimulated Brillouin Scattering: Fundamentals and Applications* (Institute of Physics Pub, Bristol, 2003).

[16] R.W. Boyd, *Nonlinear Optics*, Third edition (Academic Press, 2008).

[17] Y. Yan, E.B. Gamble, and K.A. Nelson, J. Chem. Phys. **83**, 5391 (1985).

[18] K.A. Nelson, R.J.D. Miller, D.R. Lutz, and M.D. Fayer, J. Appl. Phys. **53**, 1144 (1982).

[19] C.W. Ballmann, Z. Meng, A.J. Traverso, M.O. Scully, and V.V. Yakovlev, Optica **4**, 124 (2017).

[20] B. Krug, N. Koukourakis, J.W. Czarske, and J.W. Czarske, Opt. Express **27**, 26910 (2019).

[21] G. Antonacci, T. Beck, A. Bilenca, J. Czarske, K. Elsayad, J. Guck, K. Kim, B. Krug, F. Palombo, R. Prevedel, and G. Scarcelli, Biophys. Rev. **12**, 615 (2020).

[22] H. -N. Lin, R.J. Stoner, H.J. Maris, and J. Tauc, J. Appl. Phys. **69**, 3816 (1991).

[23] V.E. Gusev and P. Ruello, Appl. Phys. Rev. **5**, 031101 (2018).

[24] Y. Wang, D.H. Hurley, Z. Hua, T. Pezeril, S. Raetz, V.E. Gusev, V. Tournat, and M. Khafizov, Nat. Commun. **11**, 1597 (2020).

[25] X. Cui, W. Xu, Z. Xie, and Y. Wang, J. Mater. Chem. A **4**, 1908 (2016).

[26] Y. Tao, X. Wu, W. Wang, and J. Wang, J. Mater. Chem. C **3**, 1347 (2015).

[27] G. Su, V.G. Hadjiev, P.E. Loya, J. Zhang, S. Lei, S. Maharjan, P. Dong, P. M. Ajayan, J. Lou, and H. Peng, Nano Lett. **15**, 506 (2015).

[28] D. De, J. Manongdo, S. See, V. Zhang, A. Guloy, and H. Peng, Nanotechnology **24**, 025202 (2012).

[29] H.S. Song, S.L. Li, L. Gao, Y. Xu, K. Ueno, J. Tang, Y.B. Cheng, and K. Tsukagoshi, Nanoscale **5**, 9666 (2013).

[30] Y. Huang, E. Sutter, J.T. Sadowski, M. Cotlet, O.L.A. Monti, D.A. Racke, M.R. Neupane, D. Wickramaratne, R.K. Lake, B.A. Parkinson, and P. Sutter, ACS Nano **8**, 10743 (2014).

[31] T. Shibata, N. Kambe, Y. Muranushi, T. Miura, and T. Kishi, J. Phys. Appl. Phys. **23**, 719 (1990).

[32] S.K. Panda, A. Antonakos, E. Liarokapis, S. Bhattacharya, and S. Chaudhuri, Mater. Res. Bull. **42**, 576 (2007).

[33] G.A. Ermolaev, D.I. Yakubovsky, M.A. El-Sayed, M.K. Tatmyshevskiy, A.B. Mazitov, A.A. Popkova, I.M. Antropov, V.O. Bessonov, A.S. Slavich, G.I. Tselikov, I.A. Kruglov, S.M. Novikov, A.A. Vyshnevyy, A.A. Fedyanin, A.V. Arsenin, and V.S. Volkov, Nanomaterials **12**, 141 (2022).

[34] L.A. Burton, T.J. Whittles, D. Hesp, W.M. Linhart, J.M. Skelton, B. Hou, R.F. Webster, G. O'Dowd, C. Reece, D. Cherns, D.J. Fermin, T.D. Veal, V.R. Dhanak, and A. Walsh, J. Mater. Chem. A **4**, 1312 (2016).

[35] T. Sriv, K. Kim, and H. Cheong, Sci. Rep. **8**, 10194 (2018).

[36] M. Mnari, B. Cros, M. Amlouk, S. Belgacem, and D. Barjon, Can. J. Phys. **77**, 705 (2000).

[37] X. He and H. Shen, Phys. B Condens. Matter **407**, 1146 (2012).

[38] H. Wang, Y. Gao, and G. Liu, RSC Adv. **7**, 8098 (2017).

[39] Z.-Q. Zhen and H.-Y. Wang, Acta Phys. Pol. A **137**, 1095 (2020).

[40] S. Zhan, L. Zheng, Y. Xiao, and L.-D. Zhao, Chem. Mater. **32**, 10348 (2020).

[41] J.M. Skelton, L.A. Burton, A.J. Jackson, F. Oba, S.C. Parker, and A. Walsh, Phys. Chem. Chem. Phys. **19**, 12452 (2017).

[42] F. Vialla and N.D. Fatti, Nanomaterials **10**, 2543 (2020).

[43] R.S. Mitchell, Y. Fujiki, and Y. Ishizawa, Nature **247**, 537 (1974).

[44] H. Katzke, P. Tolédano, and W. Depmeier, Phys. Rev. B **69**, 134111 (2004).

[45] A.J. Smith, P.E. Meek, and W.Y. Liang, J. Phys. C Solid State Phys. **10**, 1321 (1977).

[46] P. Gicala, M. Cheng, T.S. Lott, K. Du, S.-W. Cheong, A.A. Petruk, K. Pichugin, and G. Sciaini, Appl. Phys. Lett. **118**, 264101 (2021).

[47] P.C.Y. Chang, J.G. Walker, and K.I. Hopcraft, J. Quant. Spectrosc. Radiat. Transf. **96**, 327 (2005).